# Jet Deflection via Cross winds: Laboratory Astrophysical Studies

S. .V. Lebedev[1], D. Ampleford[1], A. Ciardi[1], S.N. Bland[1], J.P. Chittenden[1], M.G. Haines[1]
And
A. Frank[2,3], E. G. Blackman[2,3], A. Cunningham[2,3]

[1] The Blackett Laboratory, Imperial College, London SW7 2BW, UK
[2] Department of Physics and Astronomy, University of Rochester, Rochester NY 14627-0171
[3] Laboratory for Laser Energetics, University of Rochester, Rochester NY 14627-0171

**Abstract:** We present new data from High Energy Density (HED) laboratory experiments designed to explore the interaction of a heavy hypersonic radiative jet with a cross wind. The jets are generated with the MAGPIE pulsed power machine where converging conical plasma flows are produced from a cylindrically symmetric array of inclined wires. Radiative hypersonic jets emerge from the convergence point. The cross wind is generated by ablation of a plastic foil via soft-X-rays from the plasma convergence region. Our experiments show that the jets are deflected by the action of the cross wind with the angle of deflection dependent on the proximity of the foil. Shocks within the jet beam are apparent in the data. Analysis of the data shows that the interaction of the jet and cross wind is collisional and therefore in the hydro-dynamic regime. MHD plasma code simulations of the experiments are able to recover the deflection behaviour seen in the experiments. We consider the astrophysical relevance of these experiments applying published models of jet deflection developed for AGN and YSOs. Fitting the observed jet deflections to quadratic trajectories predicted by these models allows us to recover a set of plasma parameters consistent with the data. We also present results of 3-D numerical simulations of jet deflection using a new astrophysical Adaptive Mesh Refinement code. These simulations show highly structured shocks occurring within the beam similar to what was observed in the experiments.

# I. INTRODUCTION

Collimated plasma beams also known as "jets" are ubiquitous in astrophysics occurring in wide variety of environments, most notably those associated with young stars (Young Stellar Objects, YSOs, Bally & Reipurth 2001) and Active Galactic Nuclei (AGN, Balsara & Norman 1992). In both environments a subset of observed bipolar jets shows a characteristic **C**-shaped symmetry. In the context of AGN such "bending jets" are known as Narrow-Angle-Tail sources (Balsara & Norman 1992). NGC 1265 is a particularly striking example of this class of objects (O'Dea and Owen 1986). Until recently only a few examples of YSO jets (also know as Herbig-Haro or HH objects) showed such bending (Bernards 5; Bally, Devine & Alten 1996, HH 30; Lopez et al 1995). Recently however Bally & Riepurth 2001b reported observations of a number of jet systems which showed bending along either a single jet in a bipolar jet system or the **C**-shaped morphological pattern for both jets. Thus it appears that the deflection of collimated plasma flows can also occur in a variety of astrophysical systems.

A number of authors have studied the cause of jet deflection and their conclusions tend to support the hypothesis that bending occurs due to the ram pressure of a "cross wind" (Begelman, Rees & Blandford, 1979, Canto & Raga 1995). In the context of AGN the crosswind is provided by the motion of the host galaxy through the IGM. NGC 1265, for example, moves at a velocity of 2200 km/s with respect to the surrounding cluster material (Chincarini & Rood (1971). For the case of YSO outflows the cross wind may be produced in two ways. All the deflected HH jets observed by Bally & Reipurth 2001b were externally illuminated, meaning they occurred close enough to massive high luminosity stars for UV fluxes to ionise the bulk of material in the jet beam. For deflected jets near the reflection nebula NGC 1333 (in the Perseus molecular cloud) Bally & Reipurth argue that, like the AGN case, rapid motion of the jet-producing stars though the star forming region produces the cross wind ram pressure effect (they suggest the stars were ejected in 3-body interactions). For deflected jets in the Orion nebula however, Bally and Reipurth argue that either a photo-ablation induced rocket effect or a cross wind from the expanding HII region produce the bending. Recent numerical simulations of the HH 505 deflected system, observed by Bally & Reipurth in Orion, show that the cross wind hypothesis is able to recover observed jet characteristics quite well (Masciadri & Raga 2001).

Analytical models of jet deflection by a cross wind *neglecting the presence of shocks* have been carried out by a Begelman, Rees & Blandford 1979 and Canto & Raga 1995. Numerical simulations of the process have been performed by a number of authors including Balasra & Norman (1992) for the extragalactic case and Lim & Raga 1998 for the YSO case. While all of these simulations demonstrated good agreement with existing analytical models, the Balsara & Norman study pointed out the importance of oblique shocks driven by the cross wind within the jet beam . Upon initial impact within the beam a "principle oblique shock" crosses the jet producing the initial deflection. Further deflection occurs through continual contact with the cross wind which produces secondary oblique shocks. Thus shock structures within the beam are important for understanding the jet morphology and the location of particle acceleration in supersonic deflected flows

While the physics of jet deflection appears well understood in terms of simulations and analysis, what is missing is direct experimental tests in a controlled setting. The advent of high energy devices used for Inertial Confinement Fusion (High Power Lasers, Fast Z Pinches) allows for direct experimental studies of astrophysical hypersonic flow problems including the effect of microphysical plasma processes. Recent laboratory studies of blast waves, shock-cloud interactions and planetary interiors have shown the promise of these kinds of scaled laboratory studies (Remington et al. 1999). High mach number jets have played an integral part in defining this new field (Logory et al. 2000, Raga et al. 2001}. Recent work by (Farley et al 1999, Shigemori et al. 2000) using high power lasers have produced radiative flows whose scalings are appropriate to make contact with astrophysical parameter regimes including radiative jets. In Lebedev et al 2001, the results of pulsed power machine studies were presented in which a conical array of 16 wires was used to generate highly collimated radiative jets. These studies were noteworthy in that they not only confirmed the utility of the pulsed power testbed for creating scalable astrophysically relevant jets they also were also able to confirm predictions of the Canto et al 1988 theory of collimation via converging conical flows and address previously unresolved issues of the stability of this collimation mechanism (Frank et al. 1996).

In this paper we present further results using the astrophysical jet testbed on the MAGPIE pulsed power device. Here we study the deflection of hypersonic jets via a cross wind produced by ablation of an irradiated target. Our goals are to explore the physics of jet deflection in a way which both expands upon, and extends, 3-D analytic studies and simulations. We seek to understand the values and limits of such studies and to further explore the utility of laboratory astrophysics experiments. The structure of our paper is as follows. In section 2 we present details of the experimental set-up. In section 3 we present the results of our experiments. Section 4 presents analysis of both the experiments and their scaling properties. In that section we present results of numerical simulations using both the plasma physics and astrophysics codes. In section 5 we present a discussion of our results, their astrophysical relevance and, finally our conclusions

## II. EXPERIMENTAL SET-UP.

**Configuration:** The schematic of the experimental set-up is shown in Fig.1. A supersonic, radiatively cooled plasma jet is produced using conical array of fine metallic wires (Lebedev et al., 2002), driven by a fast rising current (1MA, 250ns). The resistive heating of the wires rapidly converts some fraction of the wire material into a hot coronal plasma, which is then accelerated towards the array axis by the net *JxB* force. When the plasma driven off the wires reaches the array axis, a conical standing shock is formed. Such shocks are effective in redirecting the plasma flow into an axial directed flow, i.e. a jet. The high rate of radiative cooling behind this shock allows the jet to become highly collimated (Canto et al 1988). The wires act as a reservoir of material, allowing the formation and sweeping of coronal plasma into a jet to continue for the entire duration of the current pulse in the experiment. Thus our jet can be considered to be quasi-stationary. In the present experiments the array is formed by 16 (18 μm diameter) W wires. The small radius of the array is 8mm with wires inclined at an angle of $\sim 30^0$ to the array axis, and the axial length of the array is 12mm. The plasma jet formed in this configuration is hypersonic (velocity ~200 km/s, internal Mach number ~30), radiative and, most importantly, has dimensionless parameters similar to those of astrophysical (stellar) jets (Lebedev et al 2001).

The jet is driven from the wire array region into vacuum and then passes through a region with a transverse flow of plasma (the cross wind). The cross wind is produced as a result of radiative ablation of a thin plastic foil installed parallel to the axis of the jet (Fig.1). The foil is exposed to the XUV and soft X-ray radiation from the standing conical shock (the region of the jet formation) and from the wires. The intensity of this radiation is measured by an open PCD detector which has flat response in the interval 10eV – 3 keV. We find the intensity increases with time and reaches ~$10^8$ W/cm$^2$ at the time ~200 ns. Measurements with an identical detector filtered by 1.5 μm polycarbonate foil (the same material as in the foil producing the plasma flow) show that most of the radiation (>95%) is absorbed by the foil. The rate of the foil ablation and parameters of the plasma flow can be estimated from the scaling laws established in ICF research (Lindl 1995).

$$\frac{dm}{dt}[g/s/cm^2] = 10^7 (\frac{Q[W/cm^2]}{10^{15}})^{3/4} \quad (1)$$

$$V_{exhaust}[cm/s] = 1.7 \cdot 10^7 (\frac{Q[W/cm^2]}{10^{15}})^{1/8} \quad (2),$$

where Q is the incident radiation flux. For a radiation flux of $10^8$ W/cm$^2$ these formulas give a foil ablation rate of ~*55* g/s/cm$^2$ and a characteristic velocity of ablated flow leaving the foil of ~*2.3x10$^6$* cm/s. At the location where the cross wind impacts the jet, the cross winds' velocity should be somewhat higher due to expansion in vacuum. The ablation rate of the foil is increasing with time, due to an increase of the radiation flux. This allows us some degree of control over the parameters of the cross wind impacting the jet by simply changing the position of the foil with respect to the jet axis. Indeed, for a fixed moment in time (corresponding to a fixed jet dynamic age) the total amount of the ablated material and distribution with respect to the foil will be the same. However, increasing the distance between the foil and jet axis means the jet will interact with cross wind material which was ablated at earlier time (at smaller Q). As a result, the ram pressure of the cross wind ($V_{exhaust}$*dm/dt ~ $Q^{5/8}$) is smaller. In our experiments the offset of the foil from the axis was varied between 1.8 and 4.6mm. The length of the foil in the direction parallel to the jet axis was 3-5mm. Most of the experiments were performed with 1.5μm thick polycarbonate ($C_{16}H_{14}O_3$) foils. It was found that increasing the thickness of the foil or switching to CH foams instead of the foil did not change the character of the jet-wind interaction.

Diagnostics: The interaction of the jet with the plasma wind was measured using a laser probing system (λ= 532 nm, pulse duration Δt ~ 0.4 ns) with interferometer, three schlieren/shadowgraphy channels, and time-resolved XUV and soft x-ray imaging. The interferometer provided two-dimensional data on the distribution of the electron density in the jet and in the plasma flow. The schlieren diagnostic is sensitive to the gradients of the refractive index and thus highlights the regions of sharp density variations, e.g. shocks. Spatial resolution for the laser probing was ~ 0.1 mm.

## III. Experimental results

**Deflection:** Fig.3 shows the typical results of laser probing (interferometry) diagnostics of the jet propagating through the region with a cross wind and, for comparison, an image of an identical jet propagating in vacuum, without the transverse plasma flow. This figure shows that the jet remains well collimated after passing next to the foil, but is deflected away from it. The deflection depends on the position of the foil and is stronger when the foil is placed closer to the jet axis. For the image shown in Fig.3b the foil is situated 4.6mm from the axis and the jet is deflected by $\sim 4^0$. For the image in Fig.3c the foil is only 1.8mm from the axis and the deflection angle is $\sim 27^0$. The two images were obtained at the same time after the start of the current pulse and thus correspond to the same dynamic age of the jet. This also means that the same amount of material was radiatively ablated from the target by that time, and the spatial distribution of the ablated plasma with respect to the target position should be the same. However, the difference in foil positions leads to a difference in the ram pressure of the plasma wind impacting the jet, and therefore, to the difference in the degree of beam deflection. We note here that the difference in the ram pressure is mainly due to the temporal variation of the plasma flow, as we discussed in the previous section. The decrease of the ram pressure due to divergence of the flow is relatively small, at most a factor of ~2, because the dimensions of the foil are larger or comparable to the distance between the foil and jet.

**Jet and Cross Wind Parameters:** The spatial distribution of the electron density can be obtained from the interferometric images by comparing positions of the fringes with those recorded before the experiment. Fig.4 shows an interferogramm and the computed density profiles from transverse profiles of the phase shift in the probing laser beam for several axial positions. The phase shift is proportional to the line density of the electrons in the plasma ($\int n_e dl$) along the probing direction, and the shift in a single fringe corresponds to $\int n_e dl = 4.2 \times 10^{17}$ cm$^{-2}$. To transform the phase shift into electron density it is necessary to take into account the difference in the sizes of the jet (~1mm) and the plasma flowing from the foil (~1cm along the direction of the laser beam). Thus the characteristic electron densities in the wind impacting the jet are $6 \times 10^{17}$ cm$^{-3}$ and $1.7 \times 10^{18}$ cm$^{-3}$ for the images Fig.3b and 3c, respectively. The characteristic electron density of the jet in the region closest to the foil is $\sim 5 \times 10^{18}$ cm$^{-3}$. For the jet propagating in a vacuum, the density gradually decreases from the base to the tip of the jet and this decrease of density along the jet reflects the temporal evolution of the mass flow in the conical flow which produces the jet.

The impact of the wind on the jet affects the internal structure of the jet. Interferometric analysis shows appearance of asymmetries in the transverse density profiles, and a localised increase of the density gradient in the region where the jet bends. The internal structures in the jet appearing in the process of deflection are seen best on the laser schlieren images of the jet (Fig.5). The images show the presence of strong density gradients in the direction perpendicular to the jet axis. These sharp density gradients are significant in that they could be interpreted as internal shocks in the jet formed by the action of the plasma flow. In the process of deflection the diameter of the jet becomes smaller than that in the free-propagating jet, and reaches its minimum value of ~0.3 mm at the position of the sharpest change of the jet's direction. The density gradient in the jet is also strongest in this region. Above the region of jet-wind interaction, the jet expands (Fig.4, 5) to a size larger than that in the free-propagating jet. This expansion of the jet occurs in the region where the density of the wind is significantly smaller than in the interaction region, so the jet effectively propagates in vacuum. Thus the jet expansion could occur due to increase of the jet temperature after the deflection shock and / or due to reflection of the internal shock from the boundary of the jet propagating in the absence of transverse pressure balance.

As discussed above, the interferometric measurements give the electron density of the plasma. To find the mass density of both the jet and the wind it is necessary to know the corresponding ionisation states of the plasma i.e. $Z_{jet}$, $Z_{wind}$. For the CH plasma ablated from the foil by the XUV and soft X-ray radiation of the relatively low intensity, it is reasonable to assume $Z_{wind}$ = 1-2. For the jet $Z_{jet}$ is in the range 5-10, as discussed in reference (Lebedev *et al* 2002). These values of Z were used for estimates of mass density shown in Table I.

Interpretation of the observed bending of the jet will also require knowing the velocity of the cross wind. While this velocity is not measured directly, it can be estimated from the time-of-flight required for material ablated from the foil to reach the jet. Indeed, the experimental images show that the tip of the jet is deflected. This means that the cross wind has reached the axis of the system (propagated over 5mm) by the time when the jet tip was passing the foil. From the known velocity of the jet head (V=2x10$^7$ cm/s; Lebedev et al 2002) it follows that the tip of the jet was deflected ~60ns prior to the time when the images in Fig.3 were taken, i.e. at t ~250 ns after the start of the current pulse. The earliest time for the start of the foil ablation is t ~ 60ns, the time when the emission from the wires in the array starts. It is more probable, however, that the ablation of the foil starts at the time when the standing conical shock starts to form on the array axis, at t ~160 ns. At this time the radiation starts to rise sharply (Fig.2), and the radiation flux on the foil at this time reaches the level of $Q \sim 5 \times 10^7$ W/cm$^2$. Taking these two times (60ns and 160ns) as the lower and the upper limits for the starting times for the flow, we can estimate velocity of the wind as being $V_w > 2.5 \times 10^6$ cm/s and $V_w < 5.5 \times 10^6$ cm/s, respectively. Note that the value given by equation (2) is in this range (i.e. $V_w \sim V_{exhaust}$).

A summary of the characteristic parameters of the jet and the side wind in the present series of the experiments is shown in Table.I.

|  | **Density** | **Velocity (cm/s)** | **T (eV)** | **Z** |
|---|---|---|---|---|
| **Jet** | ~ $10^{-4}$ g/cm$^3$ <br> ($n_e \sim 5 \times 10^{18}$cm$^{-3}$ @ A=183, Z=10) | (10-20) x 10$^7$ | < 50eV | 5-10 |
| **Wind** | ~ $10^{-5}$ g/cm$^3$ <br> ($n_e \sim 10^{18}$ cm$^{-3}$ @ A=6, Z=1) | (2.5-5.5) x 10$^6$ | Unknown | 1-2 |

Table 1: Parameters of Jet and Cross Wind in experiments

## IV ANALYSIS

**Localisation criteria (Collisionality):** We need to check that the interaction of the plasma flow with the jet in our experimental configuration is collisional, i.e. it is similar to the hydrodynamic interaction in the astrophysical system. The condition for collisionallity is that the ratio of the particle mean free path ($\lambda$) to the jet radius be much less than unity ($\lambda/R_j \ll 1$, Ryutov et al 1999). We note that for typical YSO jet parameters ($T_j \sim .1$ eV, $n_j \sim 10^4$ cm$^{-3}$, $R_j \sim 100$ AU) the collisionality condition is achieved. For the experiment there are two aspects of the jet and the wind which need to be addressed. We need to ensure that both the jet and the wind are collisional with respect to themselves, i.e. that the mean free paths of the ions are significantly smaller than the characteristic spatial scales of the jet and cross wind. In addition we must also ensure that the interaction between the cross wind and jet is collisional in that ion collision distances are smaller that the spatial scale of the interaction region.

Despite the highly directed velocity of the jet, the temperature of both the ions and the electrons in the jet is relatively low. Estimates based on comparison with the precursor in cylindrical wire arrays and results of computer simulations indicate a temperature of $T < 50$ eV. Assuming that tungsten has $Z \sim 5$-$10$ at this temperature the upper estimate of the ion mean free path., for the typical electron densities in the jet of $n_e > 10^{18}$ cm$^{-3}$, is $\lambda_j \sim 10^{-4}$-$10^{-5}$ cm; this is significantly smaller than the jet radius ($r_j \sim 0.05$ cm). Thus the dimensionless parameter $\delta_j = \lambda_j/r_j \ll 1$ and the jet is collisional with itself on the spatial scale corresponding to the wind-jet interaction.

The cross wind is also highly collisional, even if we assume that the electron temperature is as high as 50 eV. The mean free path, for electron densities in the wind $n_e \sim 10^{18}$ cm$^{-3}$ and $Z \sim 2$, is $\lambda_w \sim 10^{-3}$ cm, which is a factor of 50 smaller than the jet radius.

The character of the impact of the side wind on the jet is determined by the m.f.p. of the wind ions interacting with the jet. For an electron density in the jet of $n_e > 5 \times 10^{18}$ cm$^{-3}$ and an ionic charge $Z > 5$, one finds $\lambda_{wj} < 2 \times 10^{-3}$ cm. Thus $\delta_{jw} = \lambda_{wj}/r_j \sim 0.04 \ll 1$ and the ions in the plasma wind will lose their momentum in collisions with the jet, therefore providing the pressure acting to deflect the jet. We conclude that the interaction of the plasma wind with the jet in our experiments can be described hydrodynamically.

**Simulation with Laboratory Codes:** We have performed two classes of simulations of our experiments. In the first class we use codes specifically developed for laboratory settings. These codes include processes such as heat conduction and cooling appropriate to tunsgsten and CH plasmas. In the second class of simulations we use astrophysical codes and attempt to explore situations with similar dimensionless parameters as the experiments. We note explicitly that the second class of simulations are not used in attempt to validate the astrophysical code, since these experiments do not yield strongly enough constrained initial and boundary conditions. Instead we use the astrophysical codes to explore similar domains of physics i.e. the effect of jet deflection from a localized cross wind. These simulations will be presented in the next section on astrophysical applications.

Simulations of the laboratory jet-wind interactions were carried out in 2D x-y slab geometry, using a two-temperature resistive magneto-hydrodynamic (MHD) code. The code also includes optically thin radiative losses, ionisation and thermal conduction (Chittenden et al 2001). The results presented here, were obtained by running the code in a purely hydrodynamic regime. From 2D MHD simulations in cylindrical symmetry of the jet formation region and the jet itself, the fluxes of mass, momentum and energy just above the conical shock were calculated and then mapped on the 2D x-y slab geometry simulations as boundary conditions. The resulting jet is both non-uniform and time-varying, with typical (axial) velocities of $v_j \sim (10\text{-}20) \times 10^6$ cm/s and densities on axis of $n_e \sim 10^{-6}\text{-}10^{-3}$ g/cm$^3$. Although the code does not allow the inclusion of more than one ion species, jet and wind material were tracked independently by solving separately the respective mass advection equations. Because the rate of radiative losses in the jet are far larger than in the wind, we set the radiative cooling in a given cell to be proportional to the amount of jet material present in that cell. Thus, when a computational cell contained exclusively wind material for example, the radiative losses were turned off. We stress that both wind and jet material were treated using the same ionisation model and equation of state.
The wind was injected perpendicular to the jet axis with a constant density of $\rho = 10^{-5}$ g cm$^{-3}$, corresponding for an average atomic mass A ~ 6, to an ion density of ~ $10^{18}$ cm$^{-3}$. Increasing the wind velocity and thus the wind ram pressure increases the jet bending angle. A wind velocity $V_w \sim 4.4 \times 10^6$ cm/s was therefore used in the simulation in order to produce a bending angle of the tip of the jet of ~10° with respect to its axis. Both the wind velocity and bending angle are consistent with the experimental values.

Mass density contour plots at t = 330 ns, for the interaction of a tungsten jet with a side wind are shown in Fig.6. Because of the time dependence of the ablation rate of the wires the jet density increases with time as $\rho_j \sim I^2$, where I is the current through the array, while the axial velocity of the jet decreases (Ciardi et al. 2002). Typical values of the ratio of wind and jet densities, and wind and jet velocities during the interaction are: $\rho_w/\rho_j \sim 3\text{-}0.3$ and $V_w/V_j \sim 0.2\text{-}0.3$ respectively. The values are characteristic of the region around the shock at the jet-wind interface, where its density is somewhat lower than its on-axis value due to the jet lateral expansion. The jet is injected in vacuum where it expands and cools, mostly by radiative losses. Where the jet and wind meet and mix, a stand-off shock develops, which persists during the whole interaction. Note that this shock is oblique and leads the initial deflection. The inclination angle $\theta$ of this shock, with respect to axial direction, decreases during the collision and depends on the relative values of the jet and wind ram pressures. We found that $\theta$ is well approximated by the simple relation

$$\theta = \arctan\left(\sqrt{\frac{\rho_w v_w^2}{\rho_j v_j^2}}\right) \qquad (3)$$

which is obtained by balancing the wind and jet ram pressures perpendicular to the shock, and disregarding both the non-uniformities along the shock and the radial expansion of the jet. In the simulations the angle $\theta$ varies between ~ 23° at 270 ns, ~

19° at 300 ns and ~ 12° at 330 ns, while the values obtained using equation (3), at the same times, are ~ 21°, ~ 16° and ~ 11° respectively. The interaction effectively bends the jet and also increases the collimation of its interacting side (Fig.6b). At 330 ns the tip of the jet is inclined by ~ 10° and moves at $V_j$ ~20.0x10$^6$ cm/s; increasingly lower inclination angles are seen further down along its axis and follow from the increasing jet ram pressure. In the interaction the wind is "dragged" along by jet (Fig.6c), gaining axial momentum. Typical axial velocities of the wind material, after the interaction, are comparable with nearby jet fluid velocities.

## V. ASTROPHYSICAL RELEVANCE

In the previous sections we have presented the results of the experiments and argued that they explore the hydrodynamic regime of jet deflection in the presence of a crosswind. In this section we show how the experiment touches directly on issues relating to astrophysics, in particular jet deflection in YSOs. In addition we use simulations with an astrophysical code to extend previous discussions of jet deflection.

**Analytic Models:** In Canto & Raga 1995 an analysis of jet deflection via a cross wind was presented for both adiabatic and isothermal jet/wind interactions. In these models the jet/wind interaction region was modelled as a "plasmon" with parabolic cross section. No distinction was made between the shocked crosswind and shocked jet material and the shape of the jet trajectory was determined via a balance between the pressure in the plasmon and the ram pressure of the cross wind. While the authors found that if the jet were to travel enough scale lengths it would, eventually be turned to flow parallel to the wind, a solution close to the initial point of jet/wind contact could be written as a quadratic in the distance x along the original jet propagation direction. When the jet is initially flowing perpendicular to the direction of the wind the solution takes the form.

$$r = \left(\frac{1}{2d}\right)(z - z_s)^2 \qquad (4)$$

where $z_s$ is the initial position at which the jet encounters the crosswind and r is the direction perpendicular to the initial flow. The principle parameter determining the deflected jet trajectory takes the form of a characteristic length scale *d*,

$$d = \sqrt{\frac{\dot{M}_j v_j^3}{\pi c^2 \rho_w v_w^2}} = \left(\frac{v_j^2}{c v_w}\right)\sqrt{\frac{\rho_j}{\rho_w}} r_j \qquad (5)$$

where *c* is the sound speed in the shocked jet.

In order to apply this expression to our experiments we could use experimental values of *c* the sound speed in the jet. These values however refer to the undisturbed jet beam before it interacts with, and is deflected by, the cross wind. Thus, in what follows we use equation 4 and 5 along with a determination of *d* from the experiments. These are combined to estimate *c* and then check for consistency with the experimentally measured Mach number in the undisturbed jet. We expect that the Mach number in the shocked jet will be lower than that in the undisturbed jet but of the same order since the shock in the jet is highly oblique.

In Figure 7 we fit the data from the experiment with three quadratics (equation 4). One of these lies directly along the jet path while the other two define minimum and maximum extent of the jets deflection based on the interferometric data. From these fits we find three values of the scale length $d$: $d_{min}$ = 5.5 cm, $d_{best}$ = 7.1 cm, $d_{max}$ = 16.6 cm. Taking the best fit and using the experimental results, 100 km/s < $v_j$ < 200 km/s, 25 km/s < $v_w$ < 55 km/s, $(\rho_j/\rho_w)$ ~ 10 and $r_j$ ~ 0.05 cm, we find limits on the Mach number in the shocked jet to be, 6 < $M$ < 26. This is consistent with the values of the Mach number, $M$ > 15 – 20, in the undisturbed jet beam inferred directly from the experimental data (Lebedev et al 2002). Such consistency not only strengthens the argument for the hydrodynamic nature of the behaviour observed in the experiments, it also demonstrates the ability of experiments to make contact with existing astrophysical theory.

**Astrophysical Simulations**: In figure 8 and 9 we show results of 3-D simulations of jet wind interactions. These simulations were carried out using AstroBEAR, a new adaptive mesh refinement code for astrophysical fluid dynamics (Poludnenko et al 2004, Varniere et al. 2004)   These simulations were designed to compare the structure of a jet interacting with a continuous cross wind (i.e. one which extends across the entire length of the jet propagation region) and one interacting with a localized cross wind (i.e one with limited spatial extent as occurs in the experiments), We note that all published astrophysical simulations of this problem utilize a continuous cross wind which fills the computational grid.  A continuous cross wind is appropriate only in cases where the jet propagation scale is smaller than the scale of any inhomogeneities in the surrounding environment which contains the cross wind. When the cross wind forms due to motion of the source this assumption may not be valid especially when the jet propagates over large distances which may take it out of its natal environment.

We note also that some of the deflected HH objects discovered by Bally & Reipurth 2001b show deflections which begin at some length down the jet beam.  These may occur due to a change in direction in the beam at some earlier point in the jets history. It is also possible, however, that background flow patterns in star forming regions may be locally heterogeneous such that an individual jet may find itself expanding into a "gust" of background wind such that the ram pressure differences will only initiate deflection at some point downwind of the jet source.

Our simulations were performed using a 3 level adaptive mesh with a maximum resolution 384 × 128 × 64.  Our simulations allowed for radiative losses using a standard cooling curve. The parameters for the simulations were similar to those found in the experiment. We use a high Mach number jet $M_j$ = 20 with a lower mach number cross wind $M_w$ = 6. In the simulation with a finite cross wind we also had to set conditions in the "ambient medium" into which both the jet and cross wind would propagate.  Parameters for the two simulations were as follows are given in Table 2.

| Jet Flow Parameters | |
|---|---|
| Computational Cells/ Radius | 10 |
| Radius | 1.496 x $10^{15}$ cm |

| Number Density | 751 cm$^{-3}$ |
|---|---|
| Temperature | 11000 K |
| Mach Number | 20 |
| Internal Mach Number | 20 |
| Wind Flow Parameters | |
| Left Edge of Wind | 3.6 Jet Radii |
| Right Edge of Wind | 23.6 Jet Radii |
| Number Density | 300 cm$^{-3}$ |
| Temperature | 11000 K |
| Mach Number | 6 |
| Internal Mach Number | 2 |
| Ambient Gas Parameters | |
| Number Density | 90 cm$^{-3}$ |
| Temperature | 99000 K |

Table 2: Astrophysical Jet-Wind Interaction Simulation Parameters

Figure 8a shows a 3-D visualization of the continuous cross wind case using two iso-density contours. Note that the jet propagation direction is along the z-axis in these simulations and the cross wind propagates upward from lower x-boundary. The deflection of the jet is apparent through its interaction with the wind. Note the ripples in the outer iso-surface defining the outer boundary of the interaction with the cross wind. These occur due cooling behind the shock. Adiabatic simulations with identical initial conditions show this boundary to be smooth. Figure 8b shows a Schlieren map of the cross-cut along the x-z plane which highlights shock waves. Here both the shock driven into the cross wind and into the jet beam are evident. The shock facing into the cross wind appears just below the jet beam at the jet injection point. The distance of this shock from the jet decreases (as one moves along the beam) until it intersects with the initial location of the beam. The shock driven into the jet becomes apparent just before this point. As one looks along the direction of the initial jet propagation we see both shocks cut diagonally across the jet beam. As the shock driven into the jet crosses the beam the direction of the beam changes and the jet is deflected upward. Thus these simulations demonstrate that shocks within the beam result from the interaction of the jet and the wind and these shocks are part of the deflection process.

In Figure 9a and 9b we show similar images from the simulation in which the cross wind is confined in extent. We initialized the simulation such that both the jet and the cross wind begin propagation at the start of the simulation. We do not show the initial interaction of the jet with the cross wind but we find that the width in z of the cross wind decreases as it propagates in the x direction due to pressure effects. Thus in Figure 9a one can see the iso-surface for the cross wind bending "inwards" from its injection point. The most interesting aspect of these simulations however is the clearly defined initiation of the jet bending at the location where the cross wind impinges on the jet beam. In Figure 9a one can make out the relatively "naked" jet beam emanating from the injection point. The Schlierien map shows the cocoon formed via the bow shock surrounding the beam near the injection point. Once the cross wind strikes the cocoon surrounding the beam a shock forms which deflects the cross wind. As this propagates diagonally the internal shock in the jet also becomes apparent and as these two shocks cross the beam, the jet begins its deflected trajectory. As the jet crosses out of the region of the cross wind we find that the shock strengths decrease. This should be compared with the fairly constant strength of the shocks seen in the continuous wind case. In spite of these differences this simulation again demonstrates that shocks within the beam result from the jet-cross wind interaction and that these shocks are a key part of the deflection process. It is noteworthy the high density features (interpreted as shocks) in the experiments only appear after the jet reaches the approximate location of the plastic foil, just as we would predict from the simulations.

## VI CONCLUSIONS

In this paper we have presented new data from High Energy Density (HED) laboratory experiments designed to explore the interaction of a heavy hypersonic radiative jet with a cross wind. Jet deflection occurs in the context of both YSO and AGN jets. In both cases astrophysical models (analytic and simulation based) have demonstrated that the most likely explanation for the deflection is the ram pressure of a cross wind. The cross wind is either due to the motion of jet source against a background medium or due to global flows from an external source such as a massive star.

In our experiments jets are generated with the MAGPIE pulsed power machine via converging conical plasma flows. The jets which are observed to emerge from the convergence point can be shown to be both radiative and hypersonic ($M > 15$) . The cross wind is generated via ablation of a plastic foil via soft-X-rays from plasma convergence region. Our experiments show that the jets will be deflected by the action of the cross wind with the angle of deflection dependent on the proximity of the foil. Shocks within the jet beam are apparent in the data.

Analysis of the data shows that the interaction of the jet and cross wind is collisional and therefore in the hydro-dynamic regime. MHD plasma code simulations of the experiments are able to recover the deflection behaviour seen in the experiments. We have considered the astrophysical relevance of these experiments applying models of jet deflection in AGN and YSOs. Fitting the observed jet deflections to quadratic trajectories predicted by these models allows us to recover a set of plasma parameters consistent with both the data and with simple shock models. We also present results of 3-D numerical simulations of jet deflection using a new astrophysical Adaptive Mesh code of jet deflection. These simulations show highly structured shocks occurring within the beam similar to what was observed in the experiments. In a real system these shocks may represent sites of particle acceleration. We conclude that our experiments provide a new window in the nature of jet deflection relevant to astrophysics. While our ability to recover the observed deflection angle with a simple analytic estimate or recover reasonable flow parameters from the deflections using a more complete model gives us confidence that our experiments are reaching relevant astrophysical behaviours, there remain uncertainties. In particular, while the structures seen in the jet in the experiments appear linked to shocks we do not see evidence for the strongest oblique shock in the wind material, which the 2-D simulations indicate, drives the initial deflection. However, we do see a strong internal shock in the jet and the narrowing of the jet after this shock. Thus the real situation may be more complex than can be recovered in either the higher resolution 2-D simulations or lower resolution 3-D simulations. This point should receive more attention in future studies.

We note that recent studies of star formation in cluster environments show that many stars may be born with significant proper motions (Bonnell, Bate, Vine 2004). For example recent studies of PV Ceph indicate it is travelling across its molecular cloud environment at speeds of order 10 km s-1 (Goodman & Arce 2003) and that its molecular outflow shows evidence for deflection. If a scenario in which high proper motions are common then it is likely that jet or molecular outflow deflection may be more common than previously expected and in many cases the signatures of deflection are difficult to see due to confusion with different sources. Thus further experimental studies which can articulate the nature of deflection for different outflow properties (such as degree of collimation) are warranted.

**Acknowledgments**: This research was sponsored by the NNSA under DOE Cooperative Agreement DE-F03-02NA00057. Support to AF was also provided at the University of Rochester by NSF grant AST-9702484 and AST-0098442, NASA grant NAG5-8428, and the Laboratory Astrophysics program at the Laboratory for Laser Energetics.

Fig 1: Schematic of the experiment, side-on (left) and end-on (right) views. The plasma jet is formed by the conical flow from the wires and is deflected by the cross wind from the radiatively ablated plastic foil.

Fig.2: Power delivered via Radiation to surface of foil which produces cross wind.

Fig 3a : Interferometric image shows well-collimated plasma jet with characteristic electron density of $10^{19}$ cm$^{-3}$ in configuration without cross wind (without foil).

Fig.3b,c: Jet deflection for two experiments with different ram pressure in the cross wind, which was achieved by variation of position of the ablated foil.

Fig 4. Interferogram of deflected jet (left) and derived from the data cross-cuts of the electron line density.

Fig.5: Laser shadowgram showing internal structure (shocks) within the jet beam generated in the process of jet deflection. Arrow at the bottom of the picture shows initial direction of jet propagation.

Fig.6: Jet deflection via 2-D simulations of experiments. The left panel shows distribution of density (jet and cross wind combined), central and right panels show density contours for the jet and the wind, respectively.

Fig 7: Interferogram of deflected jet with trajectories fit by equation 4 and 5. The two dashed lines denote minimum and maximum deflections. The solid line represents the best fit for the data. See text for details.

Figure 8: Images of density in 3-D simulations of a jet interacting with an extended cross wind at time t = 106.6 yr. The jet propagates in the z direction and the crosswind propagates in the x direction. Top image shows 2 iso-density surfaces. The bottom image shows a greyscale schlieren map of the density at a cross-cut in the x-z plane at mid-position in y.

Figure 9: Images of density in 3-D simulations of a jet interacting with a cross wind of limited extent at time t = 106.6 yr. The jet propagates in the z direction and the crosswind propagates in the x direction. Top image shows 2 iso-density surfaces. The bottom image shows a greyscale schlieren map of the density at a cross-cut in the x-z plane at mid-position in y.